\documentclass[preprint]{aastex}
\usepackage{graphics}
\usepackage{lscape}
\usepackage{epstopdf}

\makeatletter
\newcommand*{\rom}[1]{\expandafter\@slowromancap\romannumeral #1@}
\makeatother

\begin{document}

\title{Detection and Flux Density Measurements of the Millisecond Pulsar J2145$-$0750 below 100 MHz}

\author{J. Dowell}
\affil{Department of Physics and Astronomy, University of New Mexico, Albuquerque, NM  87131, USA}
\email{jdowell@unm.edu}

\author{P. S. Ray}
\affil{Space Science Division, Naval Research Laboratory, Washington, DC 20375, USA}

\author{G. B. Taylor}
\affil{Department of Physics and Astronomy, University of New Mexico, Albuquerque, NM  87131, USA}

\author{J. N. Blythe\altaffilmark{1}}
\affil{School of Physics, Georgia Institute of Technology, Atlanta, GA  3033, USA}
\altaffiltext{1}{NREIP intern at the U.S. Naval Research Laboratory}

\author{T. Clarke}
\affil{U.S. Naval Research Laboratory, Washington, DC  20375, USA}

\author{J. Craig}
\affil{Department of Physics and Astronomy, University of New Mexico, Albuquerque, NM  87131, USA}

\author{S. W. Ellingson}
\affil{Bradley Department of Electrical \& Computer Engineering, Virginia Tech, Blacksburg, VA 24061, USA}

\author{J. F. Helmboldt}
\affil{U.S. Naval Research Laboratory, Washington, DC  20375, USA}

\author{P. A. Henning}
\affil{Department of Physics and Astronomy, University of New Mexico, Albuquerque, NM  87131, USA}

\author{T. J. W. Lazio}
\affil{Jet Propulsion Laboratory, California Institute of Technology, Pasadena, CA  91109, USA}

\author{F. Schinzel}
\affil{Department of Physics and Astronomy, University of New Mexico, Albuquerque, NM  87131, USA}

\author{K. Stovall}
\affil{Center for Gravitational Wave Astronomy and Department of Physics and Astronomy, University of Texas at Brownsville, Brownsville, TX  78520, USA}

\and

\author{C. N. Wolfe}
\affil{Bradley Department of Electrical \& Computer Engineering, Virginia Tech, Blacksburg, VA 24061, USA}

\begin{abstract}
We present flux density measurements and pulse profiles for the millisecond pulsar PSR J2145$-$0750 spanning 37 to 81 MHz using data obtained from the first station of the Long Wavelength Array.  These measurements represent the lowest frequency detection of pulsed emission from a millisecond pulsar to date.   We find that the pulse profile is similar to that observed at 102 MHz.  We also find that the flux density spectrum between $\approx$40 MHz to 5 GHz is suggestive of a break and may be better fit by a model that includes spectral curvature with a rollover around 730 MHz rather than a single power law.
\end{abstract}

\keywords{pulsars: general --- pulsars: individual (PSR J2145$-$0750)}

\section{Introduction}
Millisecond pulsars (MSPs) are thought to be normal pulsars that have been spun up by accretion from a companion star \citep{Alpar82}.  These pulsars show a variety of complex pulse shapes and usually show more components than in the case of normal pulsars.  In addition, the magnetic field strengths are typically four orders of magnitude weaker than normal pulsars \citep{Kramer98}.  The flux density spectra of many MSPs can be fit by a single power law flux density spectrum down to frequencies as low as 100 MHz.  Below 100 MHz, there is evidence that at least some MSPs show a break in their spectra \citep{Erickson85,Navarro95,McConnell96,Kuzmin01}.  In addition to showing the existence of a spectral turnover, the pulse profile evolution over frequency also provides constraints on the emission mechanisms such as the height of emission region and the magnetic field configuration.  For example, PSR J2145$-$0750 is among the brightest MSPs \citep{Kuzmin01} and shows significant evolution of the pulse profile 
with frequency \citep{Kuzmin96,Kramer99}.  Above 1 GHz, the primary pulse is strongest with a weak secondary pulse that lags behind the primary by about 3 ms.  At lower frequencies, however, the strengths of the two pulses reverse and the secondary becomes dominant.  \citet{Kuzmin96} interpret this evolution in the context of quadrupole terms in the magnetic field and spatial differences between the radiating particles associated with the different components.

Beyond the pulsars themselves, low frequency observations also provide information about the intervening interstellar medium (ISM) through the sensitivity to changes in the dispersion measure (DM) and pulse scattering broadening \citep{Phillips91,PhillipsW91}.  Furthermore, the nature of the turbulence along the lines-of-sight to pulsars can also be explored through how refractive interstellar scintillation alters the observed flux density over a period of weeks \citep{Gupta93}.  This allows for a more complete understanding of the ISM that can help improve timing models used by pulsar timing arrays for the detection of low frequency gravitational waves.

The first station of the Long Wavelength Array \citep[LWA1;][]{LWA,FL}, is an ideal instrument to study pulsars at these low frequencies.  LWA1 consists of 256 dual-polarization dipole antennas that are digitally delayed and summed to form four independent beams.  Each beam has two passbands that can be tuned to a center frequency between 10 and 88 MHz, each with up to 19.6 MHz of bandwidth, that produce complex voltage data.  We have used LWA1 to study the pulsed emission from PSR J2145$-$0750 over the frequency range of 37 to 85 MHz in order to provide a better understanding of the emission mechanisms for this pulsar.


\section{Observations \& Reductions}
\label{sec:obs}
Data for PSR J2145$-$0750 were taken on two occasions using LWA1.  The first data capture is a one hour ``exploratory" data set taken on 2013 March 16 starting at 17:00 UTC with a single beam with passbands centered at 73 and 81 MHz.  The second set of observations were obtained using LWA1 on 2013 April 20 starting at 14:00 UT for two hours using three beams.  The first beam had passbands centered on 41 and 49 MHz while the second beam had passbands centered on 57 and 65 MHz.  The final beam used center frequencies of 73 and 81 MHz.  For both data sets, raw complex voltage data output at 9.8 Msamples s$^{-1}$ was recorded for each beam so that the data could be coherently dedispersed.  The useable bandwidth within each of the passbands was approximately 8 MHz.

The data were dedispersed using the coherent dedispersion functions that are part of the LWA Software Library \citep{LSL}.  These functions follow the dedispersion procedure outlined in \S5.3 of \citet{Handbook}.  The dedispersion was carried out with a DM of 9.000 pc cm$^{-3}$ \citep{Bailes94} on the LWA User Computing Facility cluster, a cluster of six nodes with 32 GB of RAM each.  This cluster was used to help manage the large amount of memory required to coherently dedisperse such a large fractional bandwidth.  For reference, the dispersion delay across the 41 MHz passband is about 10.9 seconds.  Unfortunately, the available cluster memory was not enough to process the 41 MHz passband and therefore the coherent dedispersion for this passband was performed using a computer located at the University of New Mexico containing 128 GB of RAM.

\begin{figure}
	\epsscale{0.5}
	\plotone{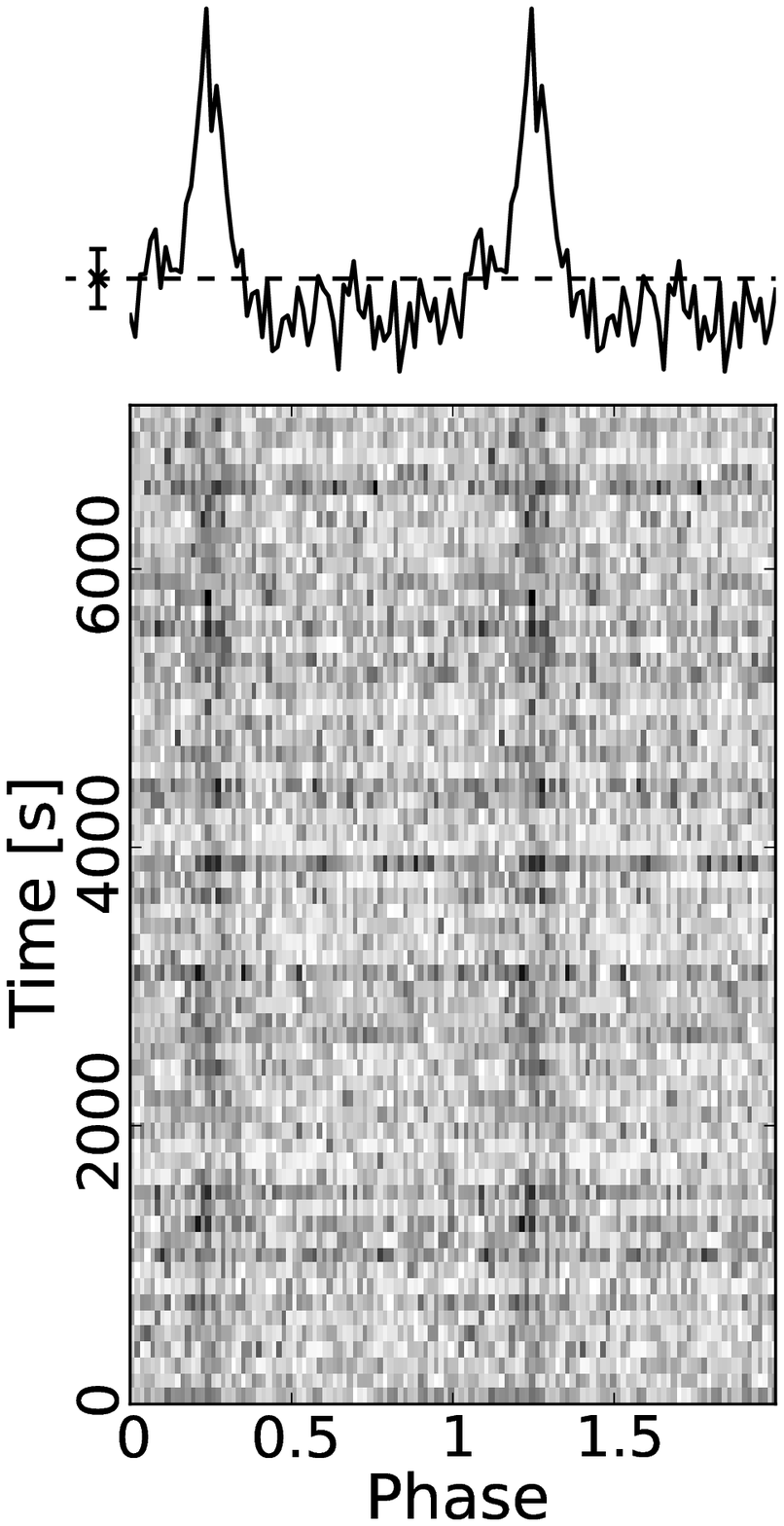}
	\caption{Coherently dedispersed average pulse profile (top) and strength as a function of time (bottom) for the two hour observation centered on 49 MHz averaged over $\sim$8 MHz of bandwidth.  The pulsar is detectable throughout the entire observation.  In the upper plot the horizontal dashed line in the average pulse profile denotes the mean value while the error bar shows the standard deviation of the profile.  \label{fig:presto}}
\end{figure}

After the individual data streams had been dedispersed, a filter bank was synthesized using a fast Fourier transform with 1,024 channels and applied to the data.  Then the two linear polarizations were summed.  The resulting spectra were written to the PSRFITS data format \citep{PSRFITS}.  This yielded spectra with a temporal resolution of $\approx$104 $\mu$s and a spectral resolution of about 9.6 kHz.  This temporal resolution is approximately one-third of the pulse width at 50\% of peak of 337 $\mu$s at 1400 MHz \citep{ATNF,Parkes} for this pulsar.  The data were then flagged for radio frequency interference and folded at the pulsar period of $\approx$16.0524 ms with PRESTO \citep{PRESTO} using the pulsar's spin and orbital parameters.  Figure \ref{fig:presto} shows an example of the detection for the passband centered on 49 MHz.

After folding, the data were calibrated using a system equivalent flux density (SEFD) estimate derived from drift scans of Virgo A taken at the same elevation as the transit of PSR J2145$-$0750 ($\approx$48$^{\circ}$).  The drift scans of the calibrator were fit with a Gaussian in order to estimate the on and off-source power.  These estimates were then combined with the flux density of \citet{Baars} to calculate the SEFD for each center frequency.  The LWA1 SEFD is dependent not only on the pointing direction but also on the local sidereal time due to the sky noise dominance of the system temperature.  To estimate the systematic uncertainty arising from this, we also measured the SEFD from drift scans of Cygnus A.  Cygnus A is not an ideal source for determining the SEFD due to its proximity to the Galactic plane and the large size of the LWA1 beam ($\sim$2$^{\circ}$ at 74 MHz) but it can be used to provide an upper limit on the SEFD and pulsar flux.

These SEFDs were then used to determine the flux density of the pulsar via:

\begin{equation}
S(\nu) = \mbox{SEFD}(\nu)\left(\frac{P_\mathrm{avg}}{P_\mathrm{off}} - 1\right) \mbox{,}
\end{equation}

\noindent where $P_\mathrm{avg}$ is the observed power averaged over the pulsar's period combined with the observed power from the sky, and $P_\mathrm{off}$ is the observed power from the sky when the pulsar is off.  For the motivation behind this approach, see \citet{LWA}.  The results from the two calibration sources were then combined with a weighted average at each frequency to yield the flux densities presented in Table \ref{tab:fluxes}.  The 1$\sigma$ uncertainties listed in this table give both the random and systematic uncertainties.  The random error arises from the uncertainty in the SEFD values and the observed sky power, and the systematic error is associated with approximating the SEFDs appropriate for PSR J2145$-$0750 at the time of the observations with those of Virgo A and Cygnus A.

In addition to our data, we also used values for the flux density of PSR J2145$-$0750 gathered from the literature for frequencies between 102 and 4850 MHz \citep{Kuzmin01,Malofeev00,Kijak97,Kramer99,Stairs99,Toscano98,Bailes94}.  The flux densities of \citet{Stairs99} and \citet{Bailes94} do not report uncertainties and for these we have adopted a 10\% random uncertainty and a 10\% systematic uncertainty.

\section{Results}
\label{sec:results}
\subsection{Dispersion Measure Variation}
\label{sec:dm}
As part of the folding process described in \S\ref{sec:obs}, a separate DM search was carried out on each passband.  Averaging the DMs estimated from each passband, we find a DM for the pulsar of 9.005$\pm$0.002 pc cm$^{-3}$.  \citet{Keith13} present a six year study of the DM of this pulsar and find noise-like fluctuations about a mean value of 8.998$\pm$0.001 pc cm$^{-3}$ (M. Keith, 2013, private communication).  The data of this study end approximately one year before our observations begin, and such a large change in the DM over that period ($\sim$0.007 pc cm$^{-3}$ yr$^{-1}$) would be unusual given the behavior shown.  A more likely explanation is that the difference in the DM arises from frequency dependent effects, such as intrinsic pulse profile evolution and scattering (see \S\ref{sec:prof}), although we cannot rule out a change in the DM based on the current data. 

\subsection{Flux Density Spectrum}
\begin{figure}
	\epsscale{1.0}
	\plotone{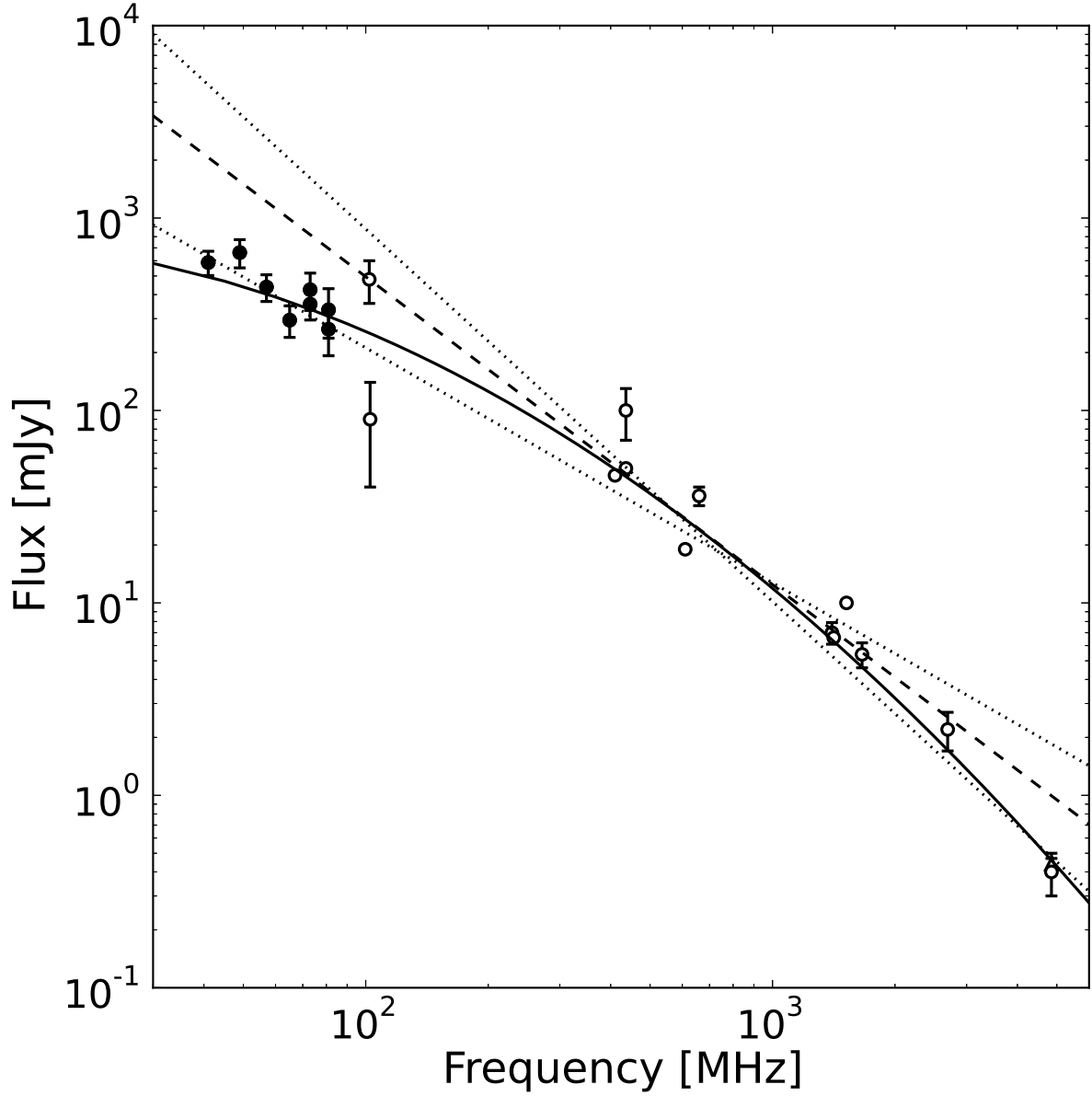}
	\caption{Flux density spectrum from 41 to 81 MHz (filled circles) for both LWA1 data sets.  The best-fit spectral curvature model with $\alpha=-1.7$ and $\beta=-0.4$ is shown as a solid line.  The best-fit power law of \citet{Kuzmin01} is shown as a dashed line while the best fit power laws to the LWA1 data and the literature data above 400 MHz are shown by dotted lines.  The spectral curvature model provides the best fit to all data.\label{fig:fluxes}}
\end{figure}

Figure \ref{fig:fluxes} plots the LWA1 flux densities along with values from the literature mentioned in \S\ref{sec:obs}.  For the LWA1 data the random and systematic components of the uncertainty have been added in quadrature for plotting.  The plot shows good agreement between the March and April measurements at 73 and 81 MHz.  Fitting a power law of $S \sim \nu^\alpha$ to our data we find the best-fit spectral index, $\alpha$, of --1.2$\pm$0.3 with a flux density of 630$\pm$80 mJy at 41 MHz.  The two closest literature values to our frequency range, \citet{Kuzmin01} at 102 MHz and \citet{Malofeev00} at 102.5 MHz, bracket our flux densities.  At low frequencies the observed flux densities of pulsars vary over time as a result of scattering and refractive scintillation by the ISM.  For example, \citet{Gupta93} made observations of nine pulsars at 74 MHz over a period of 400 days.  They found typical variations in the flux densities on the order of tens of percent with maximum variations between the faintest 
and brightest flux densities on the order of two to three.  It is possible that the discrepancy between the LWA1 data and the 102/102.5 MHz data could be a result of scintillation.  However, additional observations are needed in order to verify this.  It should also be noted that if we extrapolate the LWA1 power law to 102 MHz we find a flux density of 210 mJy, which is within 3$\sigma$ of both literature values.

\citet{Kuzmin01} report that the spectrum of PSR J2145$-$0750 is consistent with a single power law with $\alpha = -1.6\pm0.1$ over the range of 102 MHz to 5 GHz.  In contrast, our data are suggestive of a break in the power law at $\approx$730 MHz.  Indeed, if the measurements of \citet{Kijak97} and \citet{Kramer99} are taken at face value and a power law is fit using only data above 400 MHz, the spectral index steepens to a value of --1.9$\pm$0.1, supporting the idea that there is a break in the power law.  Therefore we fit the available data with a power law that includes spectral curvature of the form:

\begin{equation}
S(\nu) = S_0 \left(\frac{\nu}{\nu_r}\right)^{\alpha + \beta\times\log(\nu/\nu_r)}\mbox{,}
\end{equation}

\noindent where $\beta$ is the spectral curvature and $\nu_r$ is the rollover frequency.  We adopted a rollover frequency of 730 MHz, which is approximately where the power law fits for the LWA1 data and literature above 400 MHz cross.  Using this, we find best-fit values of $\alpha=-1.7\pm0.1$ and $\beta=-0.4\pm0.1$.  To test whether or not this more complicated spectral curvature model describes the literature data better than a single power law, we used an $F$-test to compare the $\chi^2$ values.  The resulting $F$-statistic rejects the null hypothesis that the spectral curvature model does not describe the data better than a single power law at a significance level of 0.5\%.  Thus, we suggest that the data are better described by the more complicated model.  However, additional flux density measurements in the 100 to 700 MHz range and below 40 MHz are needed to further test the robustness of this conclusion and to average out variations introduced by interstellar scintillation that may contribute to the 
observed scatter in the data toward lower frequencies.

\subsection{Pulse Profile Evolution}
\label{sec:prof}
\begin{figure}
	\epsscale{1.0}
	\plotone{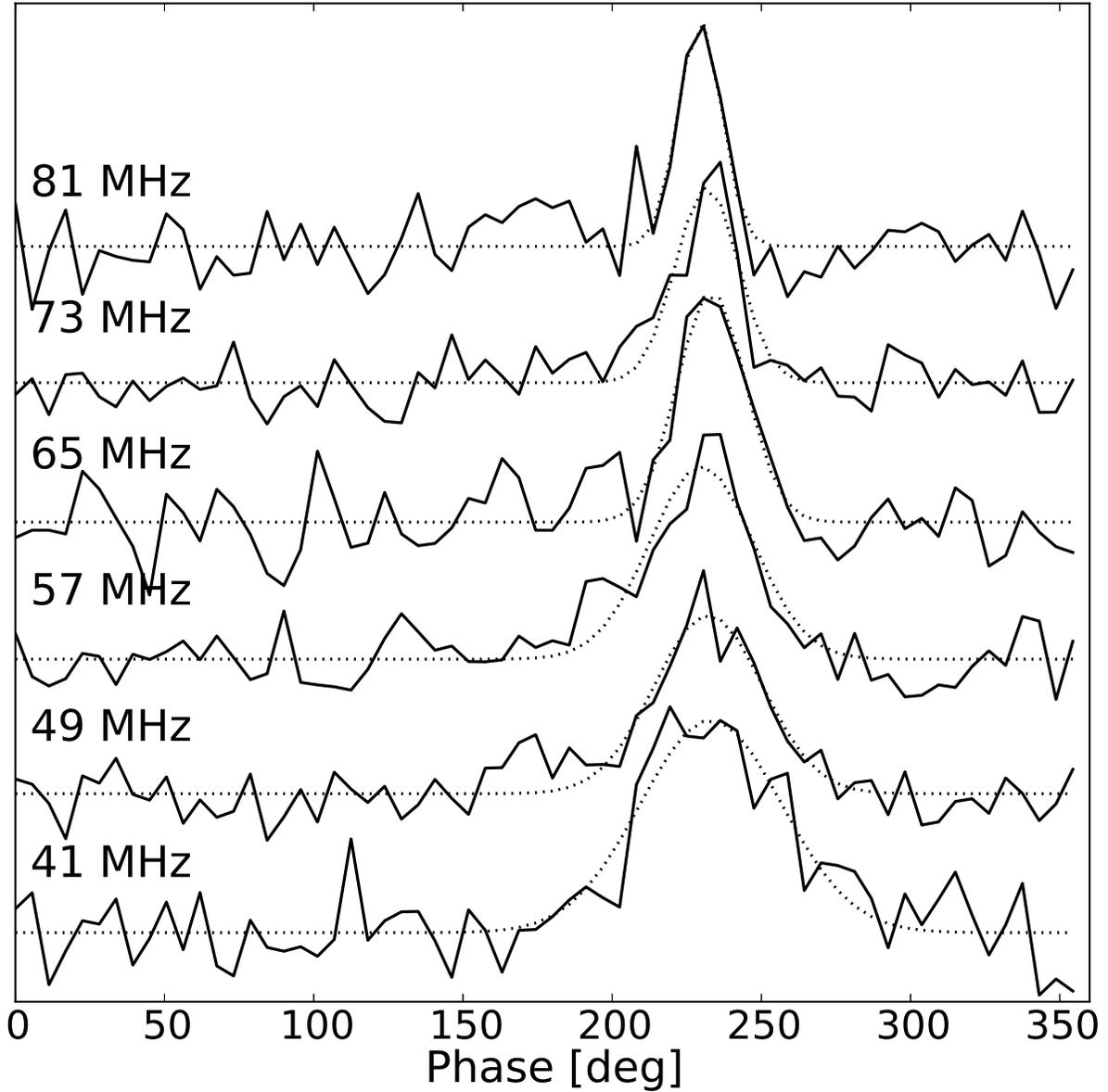}
	\caption{Pulse profiles, from bottom to top, for 41, 49, 57, 65, 73, and 81 MHz from the April observations as a function of time over one period.  Each profile has been dedispersed at our best DM value and integrated over the $\approx$8 MHz of usable bandwidth available within each passband and aligned with component \rom{2} of the 102 MHz data shown in \citet{Kuzmin96}.  Best-fit Gaussians to the pulses are indicated by dotted lines.\label{fig:profiles}}
\end{figure}

\begin{figure}
	\epsscale{1.0}
	\plotone{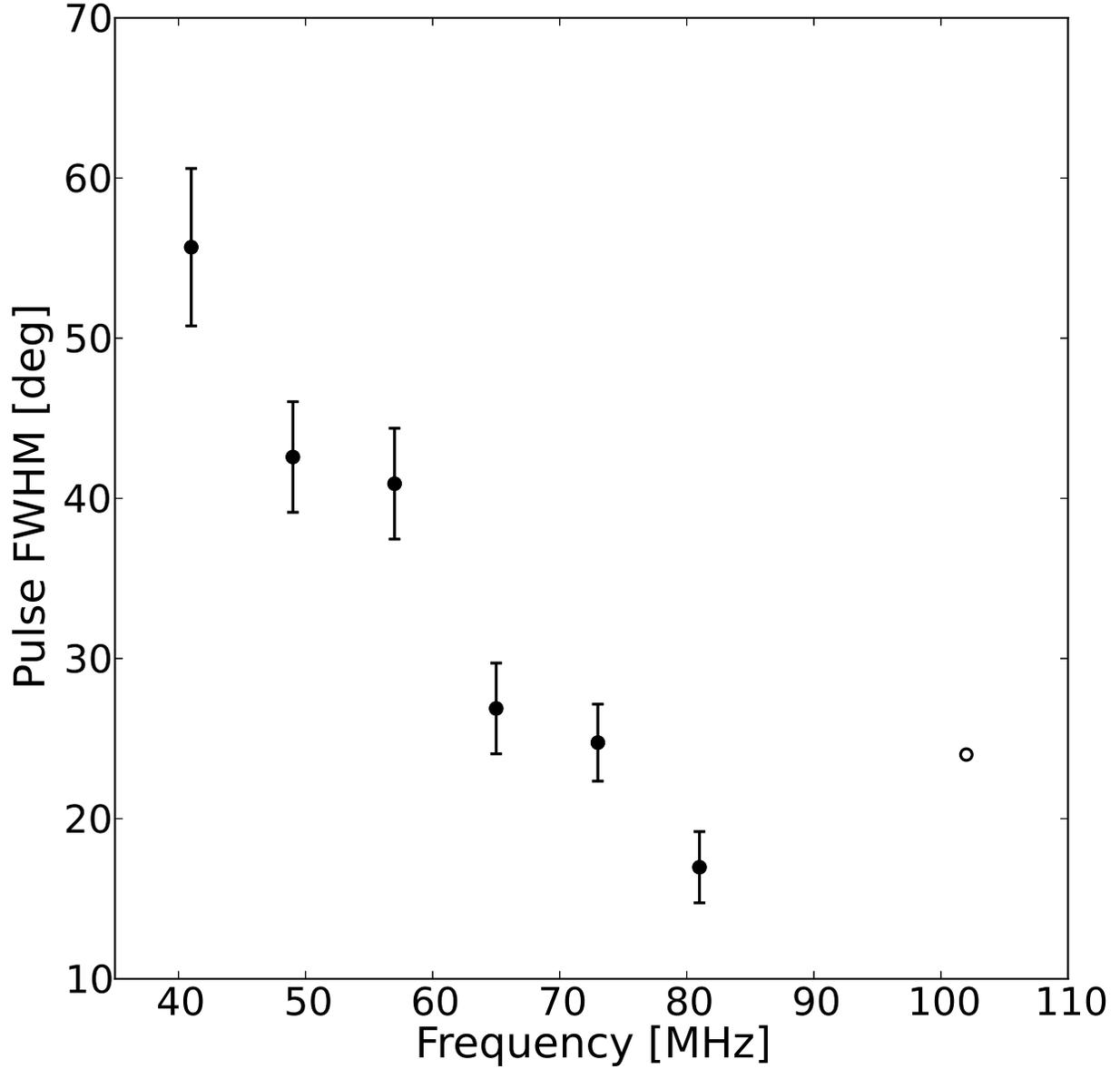}
	\caption{Evolution of the pulse full width at half maximum (filled circles) as a function of frequency for the 2013 April LWA1 data.   The 81 MHz passband pulse width is narrower than the 24 degrees reported for component \rom{2} at 102 MHz (open circle) by \citet{Kuzmin96} and the increasing pulse width toward the lower frequencies may be the result of scatter broadening of the pulse by the ISM.\label{fig:evolve}}
\end{figure}

Figure \ref{fig:profiles} shows the integrated pulse profiles over one period ($\approx$16.0524 ms) for the six center frequencies along with Gaussian fits to the profiles.  The pulses shown likely correspond to component \rom{2} in the 102 MHz profile presented in \citet{Kuzmin96} since this particular component dominates the profile at this frequency.  Accordingly, the pulses at each center frequency have been aligned with the phase of component \rom{2}.  We also tried fitting the profiles with two Gaussians but the resulting improvements were only significant at the 0.5\% level for the 49, 57, and 73 MHz passbands.  Interestingly, for the 57 and 73 MHz passbands, the second Gaussian leads the main pulse by $\sim$10 to 20 degrees which is consistent with component \rom{4} of \citet{Kuzmin96}.  However, since the majority of the profiles are not well modeled by two Gaussians we have adopted the single Gaussian fits for all six frequencies for our analysis.  We find that the full width at half maximum of the 
primary pulse in the LWA1 data increases from $\approx$17 degrees at 81 MHz to $\approx$56 degrees at 41 MHz (Figure \ref{fig:evolve}).   At the upper end of the frequency range this is narrower than the 24 degrees reported for component \rom{2} at 102 MHz.  

Besides an intrinsic increase, there are a variety of factors that could contribute to the increasing pulse width toward the lower frequencies.  For example, dedispersing at 9.000 pc cm$^{-3}$ instead of the best-fit DM value given in \S\ref{sec:dm} introduces broadening due to DM smearing.  However, this should not contribute significantly to the observed pulse width.  For the 41 MHz passband, the expected DM smearing across one channel is $\sim 0.1$ degrees (6 $\mu$s).  Another factor that contributes to the pulse width is the limited temporal resolution of the profiles presented in Figure \ref{fig:profiles} that have a resolution of approximately 6 degrees.  Finally, scatter broadening can be a factor in the pulse width evolution, particularly at low frequencies.  \citet{Johnston98} report a scintillation bandwidth of 1.48 MHz for this pulsar at a frequency of 436 MHz.  This corresponds to a pulse broadening time of $\sim$2$\times$10$^{-3}$ degrees (0.11 $\mu$s) at 436 MHz.  Since the broadening time 
scales as $\nu^{-4}$ this translates to two degrees at 81 MHz and 31 degrees at 41 MHz.  If we adopt a thin screen model for scattering and assume no evolution of the pulse profile, the observed width at 41 MHz corresponds to a width of approximately 34 degrees at 102 MHz.  This is larger that reported by \citet{Kuzmin96} and does not explain the narrower width found at 81 MHz. In addition, scattering would manifest as a significant asymmetry towards the trailing edge of the pulse profile, which is not observed.   Higher signal-to-noise ratio profiles are needed to further investigate the pulse evolution below 100 MHz.

\section{Conclusions}
\label{sec:conclusions}
We present the first flux density spectrum and pulse profiles of the MSP PSR J2145$-$0750 below 100 MHz using data obtained from the LWA1 radio telescope.  We have coherently dedispersed observations at 41, 49, 57, 65, 73, and 81 MHz, each with approximately 8 MHz of usable bandwidth and detect the pulsar over the frequency range of 37 to 85 MHz.  We find that the flux density spectrum is best fit by a model that includes a spectral curvature with $\alpha=-1.7\pm0.1$, $\beta=-0.4\pm0.1$, and a rollover frequency of 730 MHz.  However, additional measurements below 700 MHz are needed to confirm this conclusion.  We also find that the pulse profile is similar to that previously reported at 102 MHz although only component \rom{2} is detected.  Finally, these observations also serve as a proof-of-concept for future observations of MSPs with LWA1. In addition, we hope to extend observations down to lower frequencies and to higher bandwidths.

\acknowledgements{We thank the anonymous referee for their thoughtful comments.  Construction of the LWA has been supported by the Office of Naval Research under Contract N00014-07-C-0147. Support for operations and continuing development of LWA1 is provided by the National Science Foundation under grants AST-1139963 and AST-1139974 of the University Radio Observatories program.  Part of this research was carried out at the Jet Propulsion Laboratory, California Institute of Technology, under a contract with the National Aeronautics and Space Administration.}

\begin{deluxetable}{ccc}
	\tablecolumns{6}
	\tablewidth{0pc}
	\tablecaption{PSR J2145$-$0750 Flux Densities\label{tab:fluxes}}
	\tablehead{
		\colhead{Frequency} & \colhead{Date} & \colhead{$S_\nu$} \\
		\colhead{[MHz]} & ~ & \colhead{[mJy]}
	}

	\startdata
		41 & 2013 Apr 20 & 590$\pm$80$\pm$10 \\
		49 & 2013 Apr 20 & 660$\pm$70$\pm$80 \\
		57 & 2013 Apr 20 & 440$\pm$70$\pm$20 \\
		65 & 2013 Apr 20 & 290$\pm$50$\pm$20 \\
		73 & 2013 Mar 16 & 420$\pm$80$\pm$40 \\
		73 & 2013 Apr 20 & 360$\pm$50$\pm$30 \\
		81 & 2013 Mar 16 & 330$\pm$90$\pm$30 \\
		81 & 2013 Apr 20 & 260$\pm$70$\pm$30 \\
	\enddata
	
	\tablecomments{The two uncertainties listed for $S_\nu$ are for the random and systematic components, respectively.  See \S\ref{sec:obs} for details.}
\end{deluxetable}

\end{document}